\documentclass[12pt]{article}
\usepackage{graphicx}
\usepackage[cp1251]{inputenc}

\begin{document}
 \noindent {\footnotesize\it Astronomy Letters, 2018, Vol. 44, No 11, pp. 675--686.}
 \newcommand{\dif}{\textrm{d}}

 \noindent
 \begin{tabular}{llllllllllllllllllllllllllllllllllllllllllllll}
 & & & & & & & & & & & & & & & & & & & & & & & & & & & & & & & & & & & & & \\\hline\hline
 \end{tabular}

  \vskip 0.5cm
 \centerline{\bf\Large Kinematics of the Galaxy from OB Stars}
 \centerline{\bf\Large with Data from the Gaia DR2 Catalogue}
 \bigskip
 \bigskip
  \centerline
 {
 V.V. Bobylev\footnote [1]{e-mail: vbobylev@gaoran.ru} and
 A.T. Bajkova
 }
 \bigskip

  \centerline{\small\it
 Pulkovo Astronomical Observatory, Russian Academy of Sciences,}

  \centerline{\small\it
 Pulkovskoe sh. 65, St. Petersburg, 196140 Russia}
 \bigskip
 \bigskip
 \bigskip

 {
{\bf Abstract}---We have selected and analyzed a sample of OB
stars with known line-of-sight velocities determined through
ground-based observations and with trigonometric parallaxes and
proper motions from the Gaia DR2 catalogue. Some of the stars in
our sample have distance estimates made from calcium lines. A
direct comparison with the trigonometric distance scale has shown
that the calcium distance scale should be reduced by 13\%. The
following parameters of the Galactic rotation curve have been
determined from 495 OB stars with relative parallax errors less
than 30\%:
 $(U,V,W)_\odot=(8.16,11.19,8.55)\pm(0.48,0.56,0.48)$ km s$^{-1}$,
      $\Omega_0=28.92\pm0.39$ km s$^{-1}$ kpc$^{-1}$,
  $\Omega^{'}_0=-4.087\pm0.083$ km s$^{-1}$ kpc$^{-2}$ and
  $\Omega^{''}_0=0.703\pm0.067$ km s$^{-1}$ kpc$^{-3}$, where the
circular velocity of the local standard of rest is
 $V_0=231\pm5$ km s$^{-1}$ (for the adopted $R_0=8.0\pm0.15$ kpc). The parameters of
the Galactic spiral density wave have been found from the series
of radial, $V_R,$ residual tangential, $\Delta V_{circ}$, and
vertical, $W,$ velocities of OB stars by applying a periodogram
analysis. The amplitudes of the radial, tangential, and vertical
velocity perturbations are
 $f_R=7.1\pm0.3$ km s$^{-1}$,
 $f_\theta=6.5\pm0.4$ km s$^{-1}$, and
 $f_W=4.8\pm0.8$ km s$^{-1}$,
respectively; the perturbation wavelengths are
 $     \lambda_R=3.3\pm0.1$ kpc,
 $\lambda_\theta=2.3\pm0.2$ kpc, and
 $     \lambda_W=2.6\pm0.5$ kpc; and the Sun's
radial phase in the spiral density wave is
 $     (\chi_\odot)_R=-135\pm5^\circ$,
 $(\chi_\odot)_\theta=-123\pm8^\circ$, and
 $     (\chi_\odot)_W=-132\pm21^\circ$ for the adopted four-armed spiral pattern.
  }

\medskip
DOI: 10.1134/S1063773718110026

 \subsection*{INTRODUCTION}
Stars of spectral types O and B are extremely young, are
distributed over the entire Galactic disk, and are located
virtually where they were born. Such stars are members of active
star-forming regions, stellar associations, and open star clusters
(OSCs). All of this makes them very important objects in studying
the kinematics and dynamics of the Galaxy as well as its spiral
structure.

Until recently, it had been possible to use only the estimates
based on the analysis of photometric and spectroscopic data as the
distances to OB stars, because, for example, the Hipparcos
catalogue (1997) contains only $\sim$250 OB5 stars with a parallax
error $<$15\% (Maiz-Apell\'aniz 2001). The mean error of the
photometric distance estimate is $\sim$20\%. The distances to OB
stars inferred from interstellar calcium lines, whose mean error
is estimated to be 15\%, proved to be good (Megier et al. 2005,
2009; Galazutdinov et al. 2015). Important results were obtained
by various authors using the group distances to OB stars when
analyzing young OB associations and OSCs, whose mean error is
estimated to be 15--20\% (Blaha and Humphreys 1989; Zabolotskikh
et al. 2002; Mel'nik and Dambis 2009, 2017). Reliable
trigonometric parallax estimates for OB stars have been obtained
only recently.

For example, the Gaia data processing results after the first year
of its in-orbit operation were published in September 2016 (Prusti
et al. 2016; Brown et al. 2016). Its catalogue contains the
trigonometric parallaxes and proper motions of $\sim$2 billion
stars. The mean parallax error is about 0.3 mas. For some of the
stars (TGAS, Tycho--Gaia Astrometric Solution) their proper
motions were determined with a mean epoch difference of about 24
years with a mean error of about 0.06 mas yr$^{-1}$ for the stars
common to the Hipparcos catalogue (1997) and $\sim$mas yr$^{-1}$
for the remaining stars (Brown et al. 2016). Using these data,
various authors performed a number of important studies devoted to
the kinematics of stars from the solar neighborhood (Bovy 2017;
Hunt et al. 2017; Bobylev and Bajkova 2017; Vityazev et al. 2017).

The Gaia second data release, Gaia DR2, was published in April
2018 (Brown et al. 2018; Lindegren et al. 2018), while its third
data release is planned to be issued in mid-2020. The Gaia DR2
catalogue contains the trigonometric parallaxes and proper motions
of $\sim$1.7 billion stars. The derivation of their values is
based on the orbital observations performed for 22 months. The
mean errors of the trigonometric parallax and both proper motion
components in this catalogue depend on magnitude. For example, the
parallax errors lie in the range 0.02--0.04 mas for bright stars
$(G<15^m)$ and are 0.7 mas for faint stars $(G=20^m).$

It is important to note that the spatial motions of $\sim$7
million stars for which not only their highly accurate parallaxes
and proper motions, but also their line-of-sight velocities were
measured with mean errors less than 1.8 km s$^{-1}$ can be
thoroughly analyzed on the basis of Gaia DR2. However, there are
some classes of objects for which the Gaia DR2 catalogue has no
line-of-sight velocity estimates. These include, for example,
spectroscopic binary OB stars or Cepheids for which determining
the line-of-sight velocities requires special long-term
observations.

The goal of this paper is to improve the Galactic rotation
parameters and the parameters of the Galactic spiral density wave
based on OB stars using their highly accurate trigonometric
parallaxes and proper motions taken from the Gaia DR2 catalogue
with the involvement of their line-of-sight velocities measured
previously by ground-based methods.

 \subsection*{METHODS}
We know three stellar velocity components from observations: the
line-of-sight velocity $V_r$ and the two tangential velocity
components $V_l=4.74r\mu_l \cos b$ and $V_b=4.74r\mu_b$ along the
Galactic longitude $l$ and latitude $b,$ respectively, expressed
in km s$^{-1}$. Here, the coefficient 4.74 is the ratio of the
number of kilometers in an astronomical unit to the number of
seconds in a tropical year, and $r$ is the stellar heliocentric
distance in kpc. The proper motion components $\mu_l \cos b$ and
$\mu_b$ are expressed in mas yr$^{-1}$. The velocities $U,V,W$
directed along the rectangular Galactic coordinate axes are
calculated via the components $V_r, V_l, V_b:$
 \begin{equation}
 \begin{array}{lll}
 U=V_r\cos l\cos b-V_l\sin l-V_b\cos l\sin b,\\
 V=V_r\sin l\cos b+V_l\cos l-V_b\sin l\sin b,\\
 W=V_r\sin b                +V_b\cos b,
 \label{UVW}
 \end{array}
 \end{equation}
where the velocity $U$ is directed from the Sun toward the
Galactic center, $V$ is in the direction of Galactic rotation, and
$W$ is directed to the north Galactic pole. We can find two
velocities, $V_R$ directed radially away from the Galactic center
and the circular velocity $V_{circ}$ orthogonal to it pointing in
the direction of Galactic rotation, from the following equations:
 \begin{equation}
 \begin{array}{lll}
  V_{circ}= U\sin \theta+(V_0+V)\cos \theta, \\
       V_R=-U\cos \theta+(V_0+V)\sin \theta,
 \label{VRVT}
 \end{array}
 \end{equation}
where the position angle $\theta$ obeys the relation
$\tan\theta=y/(R_0-x)$, and $x,y,z$ are the rectangular
heliocentric coordinates of the star (the velocities $U,V,W$ are
directed along the corresponding $x,y,z$ axes).

To determine the parameters of the Galactic rotation curve, we use
the equations derived from Bottlinger's formulas, in which the
angular velocity $\Omega$ is expanded into a series to terms of
the second order of smallness in $r/R_0:$
\begin{equation}
 \begin{array}{lll}
 V_r=-U_\odot\cos b\cos l-V_\odot\cos b\sin l-W_\odot\sin b\\
 +R_0(R-R_0)\sin l\cos b\Omega^{'}_0\\
 +R_0(R-R_0)^2\sin l\cos b\Omega^{''}_0/2,\\
 +R_0(R-R_0)^3\sin l\cos b\Omega^{'''}_0/6,
 \label{EQ-1}
 \end{array}
 \end{equation}
 \begin{equation}
 \begin{array}{lll}
 \displaystyle
 V_l= U_\odot\sin l-V_\odot\cos l-r\Omega_0\cos b \qquad\qquad\\
 +(R-R_0)(R_0\cos l-r\cos b)\Omega^{'}_0\\
 +(R-R_0)^2(R_0\cos l-r\cos b)\Omega^{''}_0/2,\\
 +(R-R_0)^3(R_0\cos l-r\cos b)\Omega^{'''}_0/6,
 \label{EQ-2}
 \end{array}
 \end{equation}
 \begin{equation}
 \begin{array}{lll}
 V_b=U_\odot\cos l\sin b + V_\odot\sin l \sin b-W_\odot\cos b\\
    -R_0(R-R_0)\sin l\sin b\Omega^{'}_0\\
    -R_0(R-R_0)^2\sin l\sin b\Omega^{''}_0/2,\\
    -R_0(R-R_0)^3\sin l\sin b\Omega^{'''}_0/6,
 \label{EQ-3}
 \end{array}
 \end{equation}
where $R$ is the distance from the star to the Galactic rotation
axis:
 \begin{equation}
 R^2=r^2\cos^2 b-2R_0 r\cos b\cos l+R^2_0.
 \end{equation}
The quantity $\Omega_0$ is the angular velocity of Galactic
rotation at the solar distance $R_0,$ the parameters
$\Omega^{'}_0,$ $\Omega^{''}_0$ and $\Omega^{'''}_0$ are the
first, second, and third derivatives of the angular velocity,
respectively, and $V_0=|R_0\Omega_0|$.

A number of studies devoted to determining the mean distance from
the Sun to the Galactic center using its individual determinations
in the last decade by independent methods have been performed by
now. For example, $R_0=8.0\pm0.2$ kpc (Vall\'ee 2017a),
$R_0=8.4\pm0.4$ kpc (de Grijs and Bono 2017), or $R_0=8.0\pm0.15$
kpc (Camarillo et al. 2018). Based on these reviews, in this paper
we adopted $R_0=8.0\pm0.15$ kpc.

The influence of the spiral density wave in the radial $(V_R)$ and
residual tangential ($\Delta V_{circ}$) velocities is periodic
with an amplitude of $\sim$10 km s$^{-1}$. According to the linear
theory of density waves (Lin and Shu 1964), it is described by the
following relations:
 \begin{equation}
 \begin{array}{lll}
       V_R =-f_R \cos \chi,\\
 \Delta V_{circ}= f_\theta \sin\chi,
 \label{DelVRot}
 \end{array}
 \end{equation}
where
 \begin{equation}
 \chi=m[\cot(i)\ln(R/R_0)-\theta]+\chi_\odot
 \end{equation}
is the phase of the spiral density wave ($m$ is the number of
spiral arms, $i$ is the pitch angle of the spiral pattern, and
$\chi_\odot$ is the Sun’s radial phase in the spiral density
wave); $f_R$ and $f_\theta$ are the amplitudes of the radial and
tangential velocity perturbations, which are assumed to be
positive. As an analysis of the present day highly accurate data
showed, the periodicities associated with the spiral density wave
also manifest themselves in the vertical velocities $W$ (Bobylev
and Bajkova 2015; Rastorguev et al. 2017).

We apply a spectral analysis to study the periodicities in the
velocities $V_R,$ $\Delta V_{circ}$ and $W$. The wavelength
$\lambda$ (the distance between adjacent spiral arm segments
measured along the radial direction) is calculated from the
relation
\begin{equation}
 \frac{2\pi R_0}{\lambda}=m\cot(i).
 \label{a-04}
\end{equation}
Let there be a series of measured velocities $V_{R_n}$ (these can
be the radial ($V_R$), tangential ($\Delta V_{circ}$), and
vertical ($W$) velocities), $n=1,\dots,N,$ where $N$ is the number
of objects. The objective of our spectral analysis is to extract a
periodicity from the data series in accordance with the adopted
model describing a spiral density wave with parameters $f,$
$\lambda$~(or $i)$ and $\chi_\odot$.

Having taken into account the logarithmic behavior of the spiral
density wave and the position angles of the objects $\theta_n$,
our spectral (periodogram) analysis of the series of velocity
perturbations is reduced to calculating the square of the
amplitude (power spectrum) of the standard Fourier transform
(Bajkova and Bobylev 2012):
\begin{equation}
 \bar{V}_{\lambda_k} = \frac{1} {N}\sum_{n=1}^{N} V^{'}_n(R^{'}_n)
 \exp\biggl(-j\frac {2\pi R^{'}_n}{\lambda_k}\biggr),
 \label{29}
\end{equation}
where $\bar{V}_{\lambda_k}$ is the $k$th harmonic of the Fourier
transform with wavelength $\lambda_k=D/k$, $D$ is the period of
the series being analyzed,
 \begin{equation}
 \begin{array}{lll}
 R^{'}_{n}=R_0\ln(R_n/R_0),\\
 V^{'}_n(R^{'}_n)=V_n(R^{'}_n)\times\exp(jm\theta_n).
 \label{21}
 \end{array}
\end{equation}
The sought-for wavelength $\lambda$ corresponds to the peak value
of the power spectrum $S_{peak}$. The pitch angle of the spiral
density wave is derived from Eq. (9). We determine the
perturbation amplitude and phase by fitting the harmonic with the
wavelength found to the observational data. The following relation
can also be used to estimate the perturbation amplitude:
 \begin{equation}
 f_R(f_\theta, f_W)=2\times\sqrt{S_{peak}}.
 \label{Speak}
 \end{equation}
Thus, our approach consists of two steps: (i) the construction of
a smooth Galactic rotation curve and (ii) a spectral analysis of
the radial ($V_R$), residual tangential ($\Delta V_{circ}$), and
vertical ($W$) velocities. Such a method was applied by Bobylev
and Bajkova (2012, 2013, 2015, 2017) to study the kinematics of
young Galactic objects.

 \subsection*{DATA}
In this paper we use a sample of 554 OB stars. It includes:

(i) 266 single OB stars whose distances were determined previously
(Megier et al. 2005, 2009; Galazutdinov et al. 2015) from
interstellar Ca II lines;

(ii) 189 spectroscopic binary OB stars from Bobylev and Bajkova
(2013, 2015); this list includes binaries with spectral types of
the primary component no later than B2.5 and various supergiants
with luminosity classes Ia and Iab;

(iii) 99 single OB stars with spectral types no later than B2.5
that we selected previously (Bobylev and Bajkova 2013, 2015) based
on the criterion of a small (less than 10\%) relative
trigonometric parallax error in the Hipparcos catalogue (1997).

For all the listed OB stars of our sample we took the
trigonometric parallaxes and proper motion components from the
Gaia DR2 catalogue (Brown et al. 2018).

We collected the line-of-sight velocities from published sources.
This is particularly important for spectroscopic binary OB stars,
because reliable systemic line-of-sight velocities for such
binaries are determined only through long-term spectroscopic
observations and subsequent analysis of their orbital motion. The
continuously updatable SB9 bibliographic database (Pourbaix et al.
2004) is a very important source of information about the results
of an orbital analysis of spectroscopic binaries.

\begin{figure}[t]
{\begin{center}
   \includegraphics[width=0.75\textwidth]{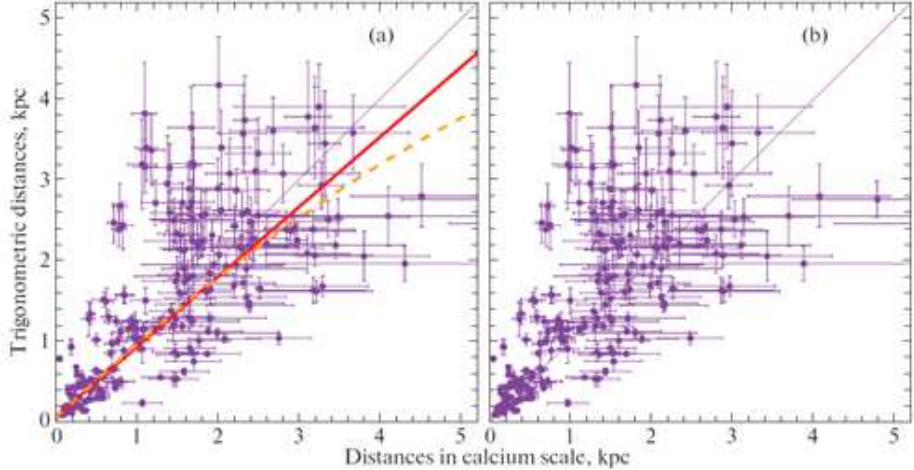}
 \caption{
(Color online) (a) Distances calculated from the Gaia DR2
trigonometric parallaxes versus distances to the stars inferred
from interstellar calcium lines; (b) the same, but the calcium
distance scale was reduced by 10\%. The thin line corresponds to
the expected dependence with a correlation coefficient equal to
one, the thick line indicates the linear dependence with
parameters (14), and the dashed line indicates the quadratic
dependence with parameters (15).
  } \label{f-dist}
\end{center}}
\end{figure}
\begin{figure}[t]
{\begin{center}
   \includegraphics[width=0.75\textwidth]{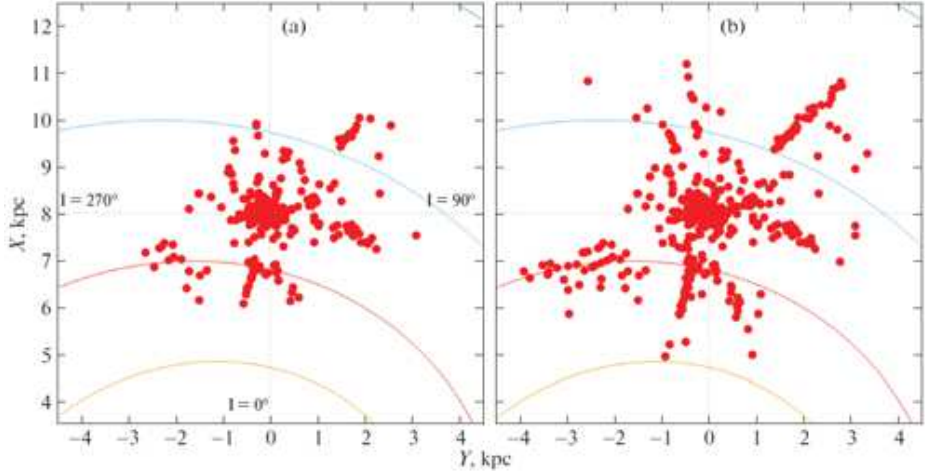}
 \caption{
(Color online) (a) Projection of the OB stars with relative
trigonometric parallax errors less than 10\% onto the Galactic
$XY$ plane; (b) the same, but with relative trigonometric parallax
errors less than 30\%. Segments of the four-armed spiral pattern
with a pitch angle of $-13^\circ$ constructed as prescribed by
Bobylev and Bajkova (2014) are shown.} \label{f-XY}
\end{center}}
\end{figure}
 \begin{table}[t]
 \caption[]{\small
The Galactic rotation parameters found from OB stars with their
proper motions and trigonometric parallaxes from the Gaia DR2
catalogue for various constraints on the relative parallax error.
The results obtained by excluding the Gould Belt stars are given
at the bottom
 }
  \begin{center}  \label{t:01}
  \small
  \begin{tabular}{|l|r|r|r|r|r|}\hline
   Parameters                 & $\sigma_\pi/\pi<10\%$ & $\sigma_\pi/\pi<20\%$ & $\sigma_\pi/\pi<30\%$ & $\sigma_\pi/\pi<100\%$  \\\hline
    $U_\odot,$    km s$^{-1}$ & $ 8.67\pm0.57$  & $ 8.52\pm0.51$  & $  8.44\pm0.52$  & $  8.54\pm0.55$  \\
    $V_\odot,$    km s$^{-1}$ & $13.28\pm0.65$  & $11.70\pm0.59$  & $ 11.59\pm0.60$  & $ 11.36\pm0.65$  \\

  $\Omega_0,$     km s$^{-1}$ kpc$^{-1}$ & $30.21\pm0.67$  & $28.45\pm0.47$   & $ 28.78\pm0.45$  & $ 28.85\pm0.43$  \\
  $\Omega^{'}_0,$ km s$^{-1}$ kpc$^{-2}$ & $-4.36\pm0.14$  & $-4.098\pm0.098$ & $-4.073\pm0.091$ & $-4.060\pm0.087$ \\
 $\Omega^{''}_0,$ km s$^{-1}$ kpc$^{-3}$ & $ 1.05\pm0.15$  & $ 0.746\pm0.076$ & $ 0.723\pm0.073$ & $ 0.710\pm0.081$ \\
$\Omega^{'''}_0,$ km s$^{-1}$ kpc$^{-4}$ &            ---  &              --- &             ---  & $-0.025\pm0.021$ \\

   $\sigma_0,$    km s$^{-1}$    &           9.68  &          10.85  &           11.33  &          12.26  \\
     $N_\star$                   &            306  &            467  &             495  &            511  \\
     $A,$ km s$^{-1}$ kpc$^{-1}$ & $ 17.43\pm0.54$ & $ 16.39\pm0.38$ &  $ 16.29\pm0.36$ & $ 16.24\pm0.35$ \\
     $B,$ km s$^{-1}$ kpc$^{-1}$ & $-12.79\pm0.86$ & $-12.06\pm0.61$ &  $-12.49\pm0.58$ & $-12.61\pm0.56$ \\
             $V_0,$ km s$^{-1}$  & $   242\pm6$    & $   228\pm6$    &  $   230\pm6$    & $   231\pm6$    \\
 \hline
    $U_\odot,$    km s$^{-1}$    & $ 6.72\pm0.97$  & $ 6.79\pm0.71$  & $  6.44\pm0.73$  & $  6.84\pm0.78$  \\
    $V_\odot,$    km s$^{-1}$    & $10.36\pm1.29$  & $ 8.78\pm0.94$  & $  8.96\pm0.95$  & $  8.91\pm1.05$  \\

  $\Omega_0,$     km s$^{-1}$ kpc$^{-1}$ & $28.92\pm0.81$  & $28.45\pm0.51$   & $ 28.69\pm0.50$  & $ 29.12\pm0.50$  \\
  $\Omega^{'}_0,$ km s$^{-1}$ kpc$^{-2}$ & $-4.15\pm0.16$  & $-4.075\pm0.102$ & $-4.042\pm0.101$ & $-4.129\pm0.099$ \\
 $\Omega^{''}_0,$ km s$^{-1}$ kpc$^{-3}$ & $ 0.82\pm0.20$  & $ 0.612\pm0.090$ & $ 0.590\pm0.088$ & $ 0.609\pm0.103$ \\
$\Omega^{'''}_0,$ km s$^{-1}$ kpc$^{-4}$ &            ---  &              --- &             ---  & $-0.006\pm0.027$ \\

   $\sigma_0,$    km s$^{-1}$    &          11.56  &          11.70  &           12.58  &          13.86  \\
     $N_\star$                   &            154  &            285  &             310  &            325  \\
     $A,$ km s$^{-1}$ kpc$^{-1}$ & $ 16.58\pm0.65$ & $ 16.30\pm0.41$ &  $ 16.17\pm0.40$ & $ 16.52\pm0.40$ \\
     $B,$ km s$^{-1}$ kpc$^{-1}$ & $-12.34\pm1.04$ & $-12.14\pm0.65$ &  $-12.52\pm0.65$ & $-12.60\pm0.63$ \\
  \hline
 \end{tabular}\end{center} \end{table}

 \subsection*{RESULTS AND DISCUSSION}
 \subsubsection*{Comparison with the Calcium Distance Scale}
We have a homogeneous set of OB stars whose distances were
determined from interstellar calcium lines at our disposal. This
distance scale is of interest for kinematic studies. Therefore, it
is important to compare these distances with the trigonometric
ones.

In Fig. 1 the distances to the OB stars calculated from the Gaia
DR2 trigonometric parallaxes are plotted against their distances
inferred from interstellar calcium lines. It can be seen from Fig.
1a that the calcium distance scale is slightly longer than the
trigonometric one. However, the effect is tangible only at
distances greater than 2 kpc. We decided to analytically check the
difference between the distance scales based on the following
relation:
 \begin{equation}
 d_{trig}=a+b\cdot d_{CaII}+c\cdot d^2_{CaII}.
 \label{correl}
 \end{equation}
A weight inversely proportional to the square of the errors,
 $p=1/(\sigma^2_{d_{trig}}+\sigma^2_{d_{CaII}})$, was assigned to each conditional
equation (13). As a result, we obtained the following solution of
the system of conditional equations (13) by the least-squares
method (LSM) based on a sample of 228 OB stars with two unknowns:
 \begin{equation}
 \begin{array}{lll}
  a=0.05\pm0.02~\hbox{kpc},\\
  b=0.87\pm0.05,
 \label{Solution-1}
 \end{array}
 \end{equation}
and with three unknowns: \begin{equation}
 \begin{array}{lll}
 a= 0.04\pm0.02~\hbox{kpc},\\
 b= 0.95\pm0.11,\\
 c=-0.042\pm0.048~\hbox{kpc$^{-1}$}.
 \label{Solution-2}
 \end{array}
 \end{equation}
Both dependences found, the linear (14) and quadratic (15) ones,
are presented in Fig. 1a.

The approach based on the solution (14) is simplest. Here the
parameter $b$ acts as a scale factor. We proceed from the
assumption that the distance scale specified by the trigonometric
parallaxes from the Gaia DR2 catalogue is most accurate. The value
of $b=0.87$ suggests that all distances to the OB stars with the
calcium distance scale should be reduced by 13\%. In the approach
based on the solution (15) we have a variable scale factor. For
example, at a distance of 3 kpc the scale factor is 0.82.

Figure 1b provides the distances to the OB stars where the calcium
distance scale was reduced by 10\%. We see that such a reduction
of the calcium distance scale makes the distribution of points on
the diagram more symmetric relative to the diagonal, although it
does not rule out completely the tail at very large ($>$4 kpc)
distances.

Based on the kinematic method, Bobylev and Bajkova (2011) showed
that the calcium distance scale is extended approximately by 20\%.
This conclusion was reached for part of the sample of OB stars
with distances exceeding 0.8 kpc considered in this paper. It is
important that the conclusion about the necessity of reducing the
calcium distance scale was corroborated by an independent method.
This allows the results of our kinematic analysis of OB stars
obtained in this paper to be compared with those obtained by us
previously with a reduction of the calcium distance scale without
taking any additional measures.

Apart from the scale factor $b$ (see Eq. (13)), the possible
systematic shift $\Delta\pi$ in the stellar parallaxes from the
Gaia DR2 catalogue with respect to an inertial frame of reference
of great interest. However, a highly accurate calibration sample
of stars that we do not have in this paper is needed for its
determination.

But Lindegren et al. (2018) have already pointed out the existence
of such a shift with $\Delta\pi=-0.029$ mas. At present, there are
several reliable distance scales a comparison with which allows,
in the opinion of their authors, the systematics in the Gaia
trigonometric parallaxes to be controlled. For example, by
comparing the parallaxes of 89 stars from the Gaia DR2 catalogue
and calibration eclipsing binary stars, Stassun and Torres (2018)
found a shift between the systems $\Delta\pi=-0.082\pm0.033$ mas.
Such a value also finds a confirmation in the works of our
authors. In particular, $\Delta\pi=-0.046\pm0.013$ mas (Riess et
al. 2018) and $\Delta\pi=-0.083\pm0.002$ mas (Zinn et al. 2018)
when analyzing Cepheids and astroseismological data, respectively.

We solved the system of equations (13) by the LSM based on a
sample of 228 OB stars, where the trigonometric distances were
corrected for the Lutz-Kelker bias. The corrections for this bias
were calculated using Eq. (28) from Rastorguev et al. (2017),
where a flat spatial distribution of stars is assumed. As a
result, we obtained the following solution with two unknowns:
 \begin{equation}
 \begin{array}{lll}
  a=0.05\pm0.02~\hbox{kpc},\\
  b=0.91\pm0.05,
 \label{Solution-11}
 \end{array}
 \end{equation}
and with three unknowns:
 \begin{equation}
 \begin{array}{lll}
 a= 0.03\pm0.02~\hbox{kpc},\\
 b= 0.99\pm0.11,\\
 c=-0.040\pm0.049~\hbox{kpc$^{-1}$}.
 \label{Solution-22}
 \end{array}
 \end{equation}
About 90\% of the OB stars in our list have small relative
trigonometric parallax errors, less than 20\%. Therefore, the
parameters found in the solutions (14), (16) and the solutions
(15), (17) differ insignificantly.

\begin{figure}[t]
{\begin{center}
   \includegraphics[width=0.55\textwidth]{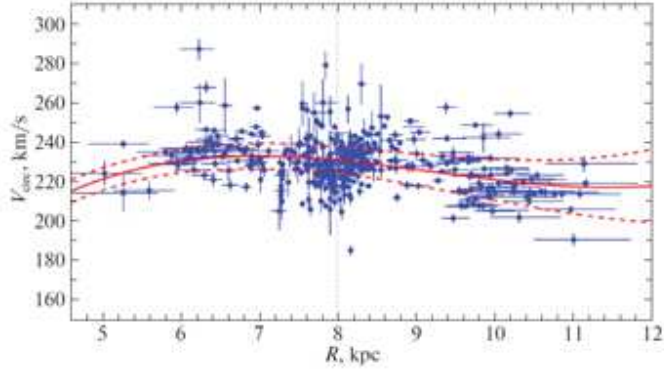}
 \caption{
(Color online) Circular velocities of OB stars with relative
trigonometric parallax errors less than 20\%, the Galactic
rotation curve constructed according to the solution (19) with the
boundaries of $1\sigma$ confidence intervals is presented, the
vertical dotted line marks the Sun’s position.
  } \label{f-circ}
\end{center}}
\end{figure}

 \subsubsection*{Galactic Rotation Curve}
The system of conditional equations (3)--(4) or (3)--(5) is solved
by the LSM with weights of the form
 $w_r=S_0/\sqrt {S_0^2+\sigma^2_{V_r}},$
 $w_l=S_0/\sqrt {S_0^2+\sigma^2_{V_l}}$ and
 $w_b=S_0/\sqrt {S_0^2+\sigma^2_{V_b}},$
where $S_0$ is the ``cosmic'' dispersion, $\sigma_{V_r},
\sigma_{V_l}, \sigma_{V_b}$ are the errors in the corresponding
observed velocities. $S_0$ is comparable to the rms residual
$\sigma_0$ (the error per unit weight) when solving the
conditional equations (3)--(5). In this paper we adopted $S_0=10$
km s$^{-1}$. For the preliminary exclusion of runaway stars we use
only one constraint on the vertical velocity, $|W|<40$ km
s$^{-1}$. The solutions are sought in two iterations with the
elimination of large residuals according to the $3\sigma$
criterion.

Figure 2 shows the distributions of our sample OB stars in
projection onto the Galactic $XY$ plane for various constraints on
the parallax errors. As can be seen from Fig. 2a, at a parallax
error less than 10\% the radius of the solar neighborhood is about
3 kpc. In this case, however, the Gould Belt and Local Arm stars
constitute the bulk, while the nearest spiral arm segments are
barely distinguishable. Our study of the distributions of OB stars
with various constraints on the parallax errors showed that to
determine the Galactic rotation and spiral structure parameters,
it is best to produce a sample with parallax errors 20--30\% (Fig.
2b).

Table 1 gives the Galactic rotation parameters found from OB stars
with various constraints on the relative parallax error. These
parameters were derived when simultaneously solving only two
equations, (3) and (4), with a different number of unknowns by the
LSM. This approach to solving such a system of equations is
commonly used in analyzing objects far from the Sun, because in
this case the contribution of Eq. (5) is negligible due to the
closeness of sin b typical for all our sample stars to zero. The
peculiar solar velocity $W$ was taken to be $W=7$ km s$^{-1}$. The
results obtained from our sample with parallax errors less than
100\% are given in the last column of the table. In this case the
equations contained six unknowns to be determined using the third
derivative in the expansion of the angular velocity of Galactic
rotation. The number of OB stars used is denoted by $N_\star$ in
the table.

Many authors use the Oort parameters to characterize the Galactic
rotation. In particular, these include the Oort constants $A$ and
$B$ that can be found from the following expressions:
\begin{equation}
 A=-0.5\Omega^{'}_0R_0,\quad
 B=-\Omega_0+A. \label{AB}
\end{equation}
It can be seen from the top part of the table that the influence
of the Gould Belt manifests itself in the sample of OB stars with
parallax errors less than 10\%. Compared to the results from other
columns, the velocities $V_\odot,$ $\Omega_0$ and $V_0$ are large
here. On the other hand, there are no significant differences
between the parameters derived with the constraints on parallax
errors less than 30\% and less than 100\%. The growth of the error
$\sigma_0$ with increasing sample radius (increasing parallax
error) is noticeable due to the influence of the proper motion and
parallax errors, because the contribution of the line-of-sight
velocity errors does not depend on the distance.

The results obtained by excluding the Gould Belt stars are given
in the bottom part of the table. For this purpose, we rejected the
stars located at distances less than 0.5 kpc. The exclusion of
such stars affected mainly the results in the first column of the
table. All of the parameters from the bottom part of the first
column now hardly differ from those in other columns. Only the
velocity $V_\odot$ changed for all the results in the bottom part
of the table. The differences in the remaining parameters are
insignificant; the random errors increased due to a considerable
number of stars.

Note the solutions obtained with six unknowns. As can be seen from
the last column of the table, the third derivative of the angular
velocity $\Omega^{'''}_0$ turns out to be small. However, when the
rotation curve is constructed from these data, the boundaries of
the confidence intervals grow very dramatically toward the edge of
the region under consideration due to the errors in
$\Omega^{'''}_0$. Therefore, below we use the solutions obtained
with the expansion of $\Omega_0$ only to the second derivative.

It is pertinent to estimate the distance $r$ at which the
contributions of the random line-of-sight velocity errors
$\sigma_{Vr}$ and the proper motion errors $\sigma_{\mu}$ become
equal. For this purpose, we use the relation
$\sigma_{Vr}=4.74\sigma_{\mu}r.$ Then, for example, for
$\sigma_{Vr}=2$ km s$^{-1}$ and  $\sigma_{\mu}=0.1$ mas we obtain
$r=4.2$ kpc, i.e., farther than 4.5--5 kpc the errors of the Gaia
DR2 proper motions dominate in the space velocities of the OB
stars. But the main conclusion here is that in the solar
neighborhood with a radius less than 4.5--5 kpc the random errors
in the space velocities $U,$ $V,$ and $W$ of the OB stars are
small (usually less than 5--6 km s$^{-1}$), which is very useful,
for example, for revealing runaway stars in this neighborhood.

Next, we simultaneously solve the system of three conditional
equations (3)--(5). Based on a sample of 495 OB stars with
relative trigonometric parallax errors less than 30\%, we found
the following kinematic parameters:
 \begin{equation}
 \label{solution-final-common}
 \begin{array}{lll}
 (U_\odot,V_\odot,W_\odot)=(8.16,11.19,8.55)\pm(0.48,0.56,0.48)~\hbox{km s$^{-1}$}, \qquad\\
      \Omega_0=~28.92\pm0.39~\hbox{km s$^{-1}$ kpc$^{-1}$},\\
  \Omega^{'}_0=-4.087\pm0.083~\hbox{km s$^{-1}$ kpc$^{-2}$},\\
 \Omega^{''}_0=~0.703\pm0.067~\hbox{km s$^{-1}$ kpc$^{-3}$}.
 \end{array}
 \end{equation}
In this solution the error per unit weight is
 $\sigma_0=10.6$ km s$^{-1}$, the Oort constants are
 $A=-16.35\pm0.33$ km s$^{-1}$ kpc$^{-1}$ and
 $B= 12.58\pm0.51$ km s$^{-1}$ kpc$^{-1}$, and the
circular velocity of the local standard of rest is
 $V_0=231\pm5$ km s$^{-1}$ for the distance
 $R_0=8.0\pm0.15$ kpc adopted in this paper.

Having analyzed the proper motions and parallaxes for a local
sample of 304 267 main-sequence stars from the Gaia DR1 catalogue,
Bovy (2017) obtained the following Oort parameters:
 $A = 15.3\pm0.5$ km s$^{-1}$ kpc$^{-1}$ and
 $B =-11.9\pm0.4$ km s$^{-1}$ kpc$^{-1}$, based on which he
estimated the angular velocity of Galactic rotation to be
 $\Omega_0=27.1\pm0.5$ km s$^{-1}$ kpc$^{-1}$ and the corresponding linear velocity to be
 $V_0 = 219\pm4$ km s$^{-1}$. Since we used considerably younger stars in
the solution (19), we obtained larger $\Omega_0$ and $V_0.$ A
different thing attracts our attention: we have three orders of
magnitude fewer stars, but obtain identical and, in several cases,
smaller errors in the parameters being determined.

It is interesting to compare the parameters (19) with the
estimates by Rastorguev et al. (2017) from 130 masers with
measured VLBI trigonometric parallaxes, where, in particular, the
two solar velocity components are
 $(U_\odot,V_\odot)=(11.40,17.23)\pm(1.33,1.09)$ km s$^{-1}$ and the
parameters of the Galactic rotation curve are
 $\Omega_0=28.93\pm0.53$ km s$^{-1}$ kpc$^{-1}$,
 $\Omega^{'}_0=-3.96\pm0.07$ km s$^{-1}$ kpc$^{-2}$ and
 $\Omega^{''}_0=0.87\pm0.03$ km s$^{-1}$ kpc$^{-3}$,
 $V_0=243\pm10$ km s$^{-1}$
(for the value of $R_0=8.40\pm0.12$ kpc found). We can see that
here there is excellent agreement in $\Omega_0$, $\Omega^{'}_0$
and $\Omega^{''}_0.$

Figure 3 provides the circular velocities of OB stars with
relative trigonometric parallax errors less than 20\%; the
Galactic rotation curve constructed according to the solution (19)
is presented. We see that the random errors of the circular
velocities for most stars are very small. At the same time, there
are a number of OB stars whose velocities deviate noticeably from
the smooth rotation curve (see also Fig. 5), i.e., they are
candidates for runaway stars.

\begin{figure}[p]
{\begin{center}
   \includegraphics[width=0.5\textwidth]{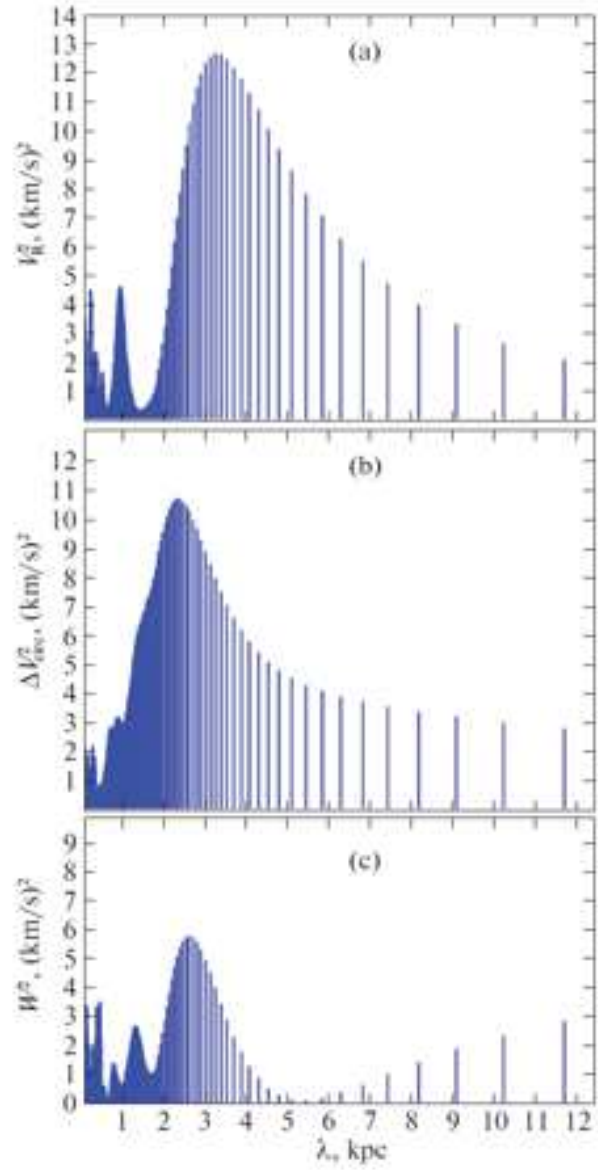}
 \caption{
(Color online) Power spectra for the radial (a), residual
tangential (b), and vertical (c) velocities of the OB stars.
  } \label{f-spectr}
\end{center}}
\end{figure}
\begin{figure}[p]
{\begin{center}
   \includegraphics[width=0.5\textwidth]{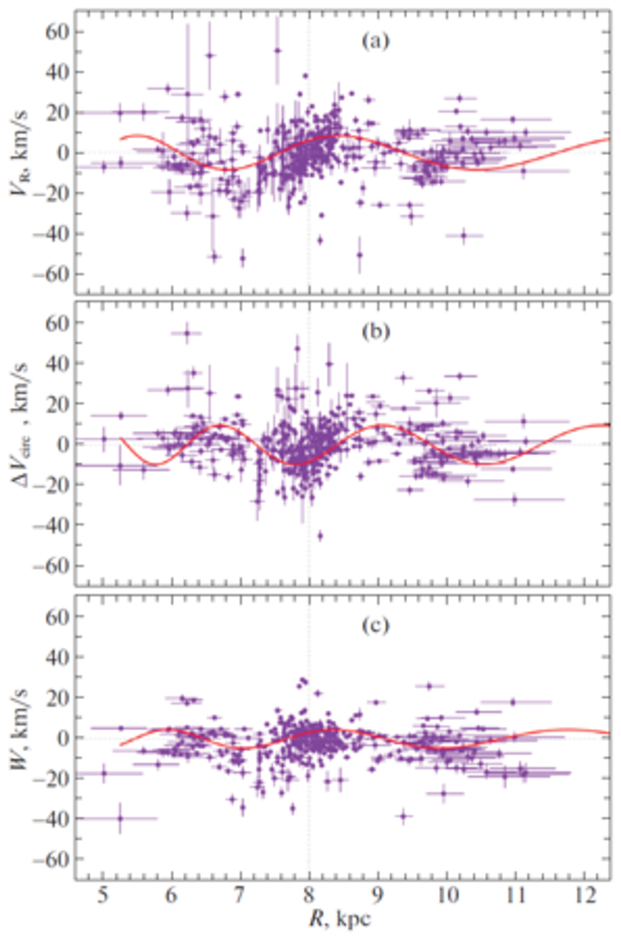}
 \caption{
(Color online) Radial (a), residual tangential (b), and vertical
(c) velocities of the OB stars versus Galactocentric distance; the
vertical dotted line marks the Sun’s position.
  } \label{f-differences}
\end{center}}
\end{figure}
\begin{figure}[p]
{\begin{center}
   \includegraphics[width=0.5\textwidth]{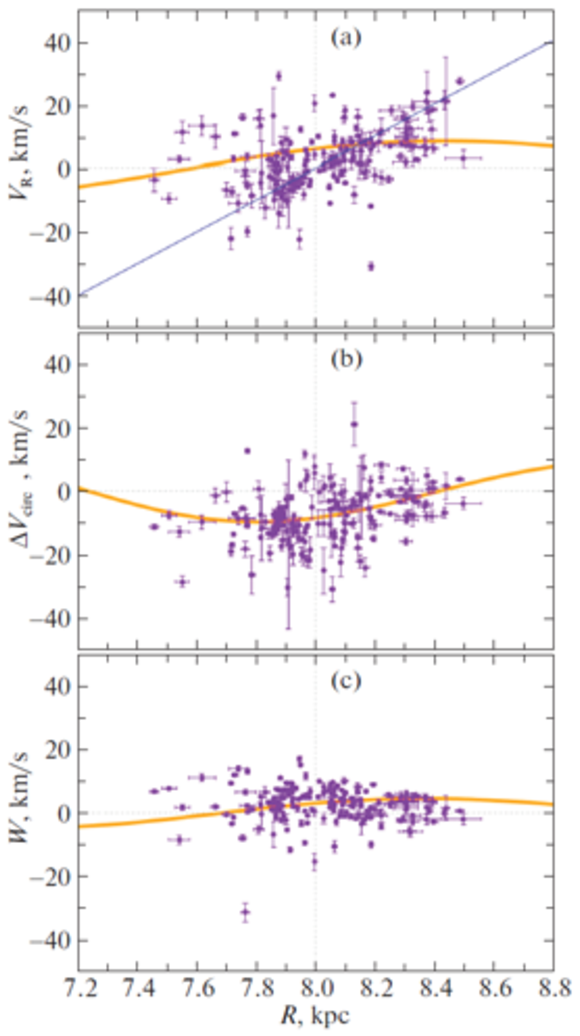}
 \caption{
(Color online) Radial (a), residual tangential (b), and vertical
(c) velocities of the Gould Belt OB stars versus Galactocentric
distance; the vertical dotted line marks the Sun’s position.
  } \label{f-GouldBelt}
\end{center}}
\end{figure}

 \subsubsection*{Parameters of the Spiral Density Wave}
For our spectral analysis we used 462 OB stars with relative
trigonometric parallax errors less than 20\%. The residual
circular velocities $\Delta V_{circ}$ were calculated from the
deviation from the Galactic rotation curve (19). In this sense the
velocities $V_R$ and $W$ are not residual, because they are
virtually independent of the Galactic rotation curve.

Based on the series of radial, $V_R,$ residual tangential, $\Delta
V_{circ}$, and vertical, $W,$ velocities of this sample, we found
the parameters of the Galactic spiral density wave with the
application of a periodogram analysis. The amplitudes of the
radial, tangential, and vertical velocity perturbations are
 $     f_R=7.1\pm0.3$ km s$^{-1}$,
 $f_\theta=6.5\pm0.4$ km s$^{-1}$,
 $     f_W=4.8\pm0.8$ km s$^{-1}$, respectively;
the perturbation wavelengths are
 $     \lambda_R=3.3\pm0.1$ kpc,
 $\lambda_\theta=2.3\pm0.2$ kpc and
 $     \lambda_W=2.6\pm0.5$ kpc
for the adopted four-armed spiral pattern ($m=4$).

Figure 4 provides the power spectra of the radial, residual
tangential, and vertical velocities for our sample of OB stars.
Good agreement in the positions of the maxima, i.e., agreement in
determining the wavelength of the spiral density wave, can be
seen. The result obtained from the tangential velocities (Fig. 4b)
should be particularly noted, because searching for the wavelength
based on such velocities previously gave a considerably lower
value, $\lambda_\theta=1.9\pm0.6$ kpc (Bobylev and Bajkova 2017).

For checking, we calculate the pitch angle of the spiral pattern
based on Eq. (9). As a result, we obtain
 $     i_R=-14.6^\circ\pm0.5^\circ$,
 $i_\theta=-10.5^\circ\pm0.7^\circ$ and
 $i_W     =-11.9^\circ\pm2.4^\circ$ from the radial, residual
tangential, and vertical velocities, respectively.

These results are in good agreement with the estimates of various
authors made in recent years. In particular, an overview of such
estimates can be found in Vall\'ee (2017b, 2017c), where the mean
values are close to $\lambda=3.1$ kpc ($R_0=8.0$ kpc), $m=4,$ and
$i=-13^\circ$.

Figure 5 provides the radial, residual tangential, and vertical
velocities of the OB stars with the inscribed periodic dependences
whose parameters were determined through a spectral analysis.
Based on this figure, we determine the Sun's radial phase in the
spiral density wave measured from the center of the
Carina.Sagittarius arm ($R\sim7$ kpc). As a result, we find
 $     (\chi_\odot)_R=-135^\circ\pm5^\circ,$
 $(\chi_\odot)_\theta=-123^\circ\pm8^\circ$ and
 $     (\chi_\odot)_W=-132^\circ\pm21^\circ$
from the radial, residual tangential, and vertical velocities,
respectively.

 \subsubsection*{On the Influence of the Gould Belt}
To study the kinematic peculiarities of the Gould Belt, from our
sample of OB stars we took stars with relative trigonometric
parallax errors less than 15\% from the solar neighborhood of
radius $r=0.6$ kpc. The produced sample contains 185 stars. The
radial, residual tangential, and vertical velocities of these
stars are given in Fig. 6. Just as in Fig. 5, the thick lines
indicate the periodic dependences whose parameters were determined
through a spectral analysis.

The thin line in Fig. 6a is given as an illustration. When
analyzing our sample of OB stars with photometric distances the
analogous velocity from stars close to the Sun was
$dU/dR\approx40$ km s$^{-1}$ kpc$^{-1}$ (Fig. 4 in Bobylev and
Bajkova (2013)).

A crowding of points in the R range 7.8--7.9 kpc is clearly seen
in all velocities of Fig. 6--in the radial, tangential and
vertical ones. Analysis of their coordinates showed that these
stars belong to the Scorpius–Centaurus association. The grouping
manifests itself particularly clearly in the radial (Fig. 6a)
velocities--the stars are aligned in a chain located at a
considerable slope (approximately twice the slope of the thin line
in the figure) to the horizontal axis. The interpretation here is
the same as that for the Gould Belt--this is the effect of
intrinsic expansion of the Scorpius–Centaurus association. Judging
by the slope, the expansion coefficient of this grouping of stars
is twice that for the Gould Belt.

Estimating accurate values for the expansion coefficient requires
determining the position of the kinematic center; it is necessary
to subtract the contributions of the Galactic differential
rotation and the spiral density wave. A larger number of stars
should also be involved. We are planning to perform such a work
later.

Excluding $\sim$150 candidates for membership in the Gould Belt
has virtually no effect on the determination of the parameters of
the Galactic rotation curve (see Table 1). However, the parameters
of the spiral density wave being determined change in this
approach. After such an exclusion we found the following
parameters: the perturbation amplitudes
 $     f_R=4.7\pm0.7$ km s$^{-1}$,
 $f_\theta=4.7\pm0.8$ km s$^{-1}$ and
 $     f_W=5.4\pm1.0$ km s$^{-1}$,
the perturbation wavelengths
 $     \lambda_R=3.0\pm0.2$ kpc,
 $\lambda_\theta=2.8\pm0.2$ kpc and
 $     \lambda_W=2.8\pm0.6$ kpc,
and the Sun's radial phase in the spiral density wave $
(\chi_\odot)_R=-180\pm8^\circ$,
 $(\chi_\odot)_\theta=-156\pm14^\circ$ and
 $     (\chi_\odot)_W=-120\pm25^\circ,$ respectively.
Compared to the previous case, where all OB stars were used, here
the agreement in determining the wavelength $\lambda$ improved,
the perturbation amplitudes $f_R$ and $f_\theta$ decreased, the
agreement in the estimates of the Sun's phase  $(\chi_\odot)_R$
and $(\chi_\odot)_\theta$, became poorer, while for the vertical
velocities, on the contrary, all parameters improved, although
their random errors slightly increased.

 \subsection*{CONCLUSIONS}
We considered a sample of 554 OB stars. The line-of-sight
velocities of these stars were determined through ground-based
observations, while their trigonometric parallaxes and proper
motions were taken from the Gaia DR2 catalogue (Brown et al.
2018).

Based on OB stars with their distances estimated from interstellar
calcium lines, we made a comparison with the distance scale
specified by the trigonometric parallaxes from the Gaia DR2
catalogue. As a result, our previous conclusion about the
necessity of a slight (less than 20\%) reduction in the calcium
distance scale was corroborated by an independent method.

The parameters of the Galactic rotation curve (19) were determined
from OB stars with reliable trigonometric parallax estimates (with
relative parallax errors less than 30\%). The derived parameters
of this rotation curve were used to form the residual velocities.

The parameters of the Galactic spiral density wave were found from
the series of radial, $V_R,$ residual tangential, $\Delta
V_{circ}$, and vertical, $W,$ velocities of OB stars by applying a
periodogram analysis. OB stars with relative parallax errors less
than 20\% were used for such an analysis. The amplitudes of the
radial, tangential, and vertical velocity perturbations are
 $     f_R=7.1\pm0.3$~km s$^{-1},$
 $f_\theta=6.5\pm0.4$~km s$^{-1},$ and
 $     f_W=4.8\pm0.8$~km s$^{-1},$ respectively; the perturbation wavelengths are
 $     \lambda_R=3.3\pm0.1$~kpc,
 $\lambda_\theta=2.3\pm0.2$~kpc, and
 $     \lambda_W=2.6\pm0.5$~kpc, adopted four-armed spiral pattern ($m=4$).
 The Sun's phase in the spiral density wave is
 $     (\chi_\odot)_R=-135^\circ\pm5^\circ,$
 $(\chi_\odot)_\theta=-123^\circ\pm8^\circ,$ and
 $     (\chi_\odot)_W=-132^\circ\pm21^\circ$.

We illustrated the influence of kinematic peculiarities of the
Gould Belt stars. For this purpose, we considered stars with
relative trigonometric parallax errors less than 15\% from the
solar neighborhood of radius $r=0.6$ kpc. The well-known expansion
of the entire complex is clearly seen in the radial velocities.
Furthermore, a fine structure manifested itself for the first time
in the radial velocities of nearby OB stars. In particular, we see
an intrinsic expansion of the stars belonging to the
Scorpius–Centaurus association.

Excluding the Gould Belt stars hardly affects the determination of
the parameters of the Galactic rotation curve. However, their
exclusion has an effect when determining the parameters of the
spiral density wave. After their exclusion we obtained the
following estimates: the perturbation amplitudes
 $     f_R=4.7\pm0.7$~km s$^{-1},$
 $f_\theta=4.7\pm0.8$~km s$^{-1},$ and
 $     f_W=5.4\pm1.0$~km s$^{-1},$
 the perturbation wavelengths
 $     \lambda_R=3.0\pm0.2$~kpc,
 $\lambda_\theta=2.8\pm0.2$~kpc, and
 $     \lambda_W=2.8\pm0.6$~kpc,
 and the Sun's radial phase in the spiral density wave
 $     (\chi_\odot)_R=-180\pm8^\circ$,
 $(\chi_\odot)_\theta=-156\pm14^\circ,$ and
 $     (\chi_\odot)_W=-120\pm25^\circ,$ respectively.

 \subsubsection*{ACKNOWLEDGMENTS}
We are grateful to the referee for useful remarks that contributed
to an improvement of the paper. This work was supported by the
Basic Research Program P--28 of the Presidium of the Russian
Academy of Sciences, the ``Cosmos: Studies of Fundamental
Processes and Their Interrelations'' Subprogram.

 \bigskip
 \bigskip\medskip{\bf REFERENCES}
{\small

1. A. T. Bajkova and V. V. Bobylev, Astron. Lett. 38, 549 (2012).

2. C. Blaha and R. M. Humphreys, Astrophys. J. 98, 1598 (1989).

3. V. V. Bobylev and A. T. Bajkova, Astron. Lett. 37, 526 (2011).

4. V. V. Bobylev and A. T. Bajkova, Astron. Lett. 38, 638 (2012).

5. V. V. Bobylev and A. T. Bajkova, Astron. Lett. 39, 532 (2013).

6. V. V. Bobylev and A. T. Bajkova, Mon. Not. R. Astron. Soc. 437,
1549 (2014).

7. V. V. Bobylev and A. T. Bajkova, Mon. Not. R. Astron. Soc. 447,
L50 (2015).

8. V. V. Bobylev and A. T. Bajkova, Astron. Lett. 41, 473 (2015).

9. V. V. Bobylev and A. T. Bajkova, Astron. Lett. 43, 159 (2017).

10. J. Bovy, Mon. Not. R. Astron. Soc. 468, L63 (2017).

11. A.G.A. Brown, A. Vallenari, T. Prusti, J. de Bruijne, F.
Mignard, R. Drimmel, C. Babusiaux, C.A.L. Bailer-Jones, et al.
(GAIA Collab.), Astron. Astrophys. 595, A2 (2016).

12. A.G.A. Brown, A. Vallenari, T. Prusti, J. de Bruijne, C.
Babusiaux, C.A.L. Bailer-Jones, M. Biermann, D. W. Evans, et al.
(GAIA Collab.), arXiv: 1804.09365 (2018).

13. T. Camarillo, M. Varun, M. Tyler, and R. Bharat, Publ. Astron.
Soc. Pacif. 130, 4101 (2018).

14. G. A. Galazutdinov, A. Strobel, F.A. Musaev, A. Bondar, and J.
Kre\l owski, Publ. Astron. Soc. Pacif. 127, 126 (2015).

15. R. de Grijs and G. Bono, Astrophys. J. Suppl. Ser. 232, 22
(2017).

16. J. A. S. Hunt, D. Kawata, G. Monari, R.J.J. Grand, B. Famaey,
and A. Siebert, Mon. Not. R. Astron. Soc. 467, 21 (2017).

17. C. C. Lin and F. H. Shu, Astrophys. J. 140, 646 (1964).

18. L. Lindegren, J. Hernandez, A. Bombrun, S. Klioner, U.
Bastian, M. Ramos-Lerate, A. de Torres, H. Steidelmuller, et al.
(GAIA Collab.), arXiv: 1804.09366 (2018).

19. J. Maiz-Apell\'aniz, Mon. Not. R. Astron. Soc. 121, 2737
(2001).

20. A. Megier, A. Strobel, A. Bondar, F. A. Musaev, I. Han, J.
Kre\l owski, and G. A. Galazutdinov, Astrophys. J. 634, 451
(2005).

21. A. Megier, A. Strobel, G. A. Galazutdinov, and J. Kre\l owski,
Astron. Astrophys. 507, 833 (2009).

22. A. M. Mel’nik and A. K. Dambis, Mon. Not. R. Astron. Soc. 400,
518 (2009).

23. A. M. Mel’nik and A. K. Dambis, Mon. Not. R. Astron. Soc. 472,
3887 (2017).

24. D. Pourbaix, A. A. Tokovinin, A. H. Batten, F. C. Fekel, W. I.
Hartkopf, H. Levato, N. I. Morrell, G. Torres, and S. Udry,
Astron. Astrophys. 424, 727 (2004).

25. T. Prusti, J.H. J. de Bruijne, A.G.A. Brown, A. Vallenari, C.
Babusiaux, C.A.L. Bailer-Jones, U. Bastian, M. Biermann, et al.
(GAIA Collab.), Astron. Astrophys. 595, A1 (2016).

26. A. S. Rastorguev, M. V. Zabolotskikh, A. K. Dambis, N. D.
Utkin, V. V. Bobylev, and A. T. Bajkova, Astrophys. Bull. 72, 122
(2017).

27. A. G. Riess, S. Casertano, W. Yuan, L. Macri, B. Bucciarelli,
M. G. Lattanzi, J.W. MacKenty, J. B. Bowers, et al., Astrophys. J.
{\bf 861}, 126 (2018).

28. K. G. Stassun and G. Torres, Astrophys. J. {\bf 862}, 61
(2018).

29. The HIPPARCOS and Tycho Catalogues, ESA SP--1200 (1997).

30. J. P. Vall\'ee, Astrophys. Space Sci. 362, 79 (2017a).

31. J. P. Vall\'ee, Astron. Rev. 132, 113 (2017b).

32. J. P. Vall\'ee, New Astron. Rev. 79, 49 (2017c).

33. V.V. Vityazev, A.S. Tsvetkov, V.V. Bobylev, and A.T. Bajkova,
Astrophysics 60, 462 (2017).

34. M. V. Zabolotskikh, A. S. Rastorguev, and A. K. Dambis,
Astron. Lett. 28, 454 (2002).

35. J. C. Zinn, M. H. Pinsonneault, D. Huber, and D. Stello,
arXiv: 1805.02650 (2018).

 }

  \end{document}